# Hole mediated ferromagnetism in Cu-doped ZnO thin films

Abhinav Pratap Singh[1], B.-G. Park[1], Ik-Jae Lee[1], Kyu Joon Lee[2], Myung-Hwa Jung[2], Jinhee Kim[3], and J.-Y. Kim[1, *]

[1]*Pohang Accelerator Laboratory, Pohang University of Science and Technology, Pohang 790-784, Korea*

[2]*Department of Physics, Sogang University, Seoul 121-742, Korea*

[3]*Electronic Devices Group, Korea Research Institute of Standards and Science, Yusung, P.O. Box 102, Taejon 305-600, Korea*

We report the successful synthesis of ZnO:Cu thin films doped with holes, resulting in room temperature ferromagnetism. Hole doping is achieved by As-diffusion from the GaAs substrate into ZnO films, assisted by thermal annealing. The As-diffusion is probed with the help of x-ray absorption spectra collected at the As $K$-edge which show enhanced signature of diffusion in the annealed samples. Introduction of holes, due to the As doping, in ZnO films is further evidenced by the Cu $L_{3,2}$-edge spectra. XMCD and magnetic measurements show that the ferromagnetic interaction between doped Cu ions is enhanced after hole doping.

[*] Electronic mail: masson@postech.ac.kr

There has been a great deal of enthusiasm among the scientific community regarding the successful synthesis of magnetic semiconductors which can act as the spin polarizers and be integrated with the present day semiconductor-based electronics. This could lead to the realization of semiconductor Spintronics promising novel applications[1]. Dietl *et al.*, used Zener's mean field model to quantitatively describe the properties of GaMnAs and ZnMnTe[2]. In the process they also predicted the possibility of above room temperature ferromagnetism in 5% Mn-doped $p$-type ZnO where it is stabilized by the holes. Recently Herng *et al.*, reported room temperature ferromagnetism in Cu-doped ZnO and suggested indirect exchange interaction among the localized Cu-moment mediated by electrons[3]. However, the exchange interaction between the localized magnetic moment and the electrons in the conduction band is weaker as compared their interaction with the holes in valence band[4]. There are reports on the successful synthesis of various transition metal doped ZnO thin films; but, the charge carriers in most of the reported work are due to the $n$-type background present in ZnO due to various defects. Thus, realization of the original prediction of Dietl *et al.* has eluded the scientific community due to the difficulty in achieving $p$-type ZnO. Recently, As-doped ZnO system has emerged as a promising candidate to prepare $p$-type ZnO[5,6].

In this letter, we report the successful preparation of the ZnO:Cu thin films which are ferromagnetic at room temperature. The ferromagnetism in the system is stabilized by the introduction of holes due to As doping. Arsenic is diffused from the GaAs substrate into the Cu-doped ZnO films. Number of holes in the films is increased by thermal annealing and the effects are observed with the help of x-ray absorption spectroscopy (XAS) measurements performed at the As $K$-edge and Cu $L_{3,2}$-edge. Superconducting quantum interference device

(SQUID) measurements showed that the films are ferromagnetic at room temperature and x-ray magnetic circular dichroism (XMCD) measurements provide a clear evidence for the ferromagnetism which is due to the exchange interaction among the localized Cu moments mediated by the holes doped in the system.

Thin films of 5% Cu-doped ZnO were deposited on the *a*-axis oriented single-crystal substrates of cubic GaAs using rf-sputtering deposition method. Prior to deposition, the substrates were etched with 10% ammonium hydroxide solution for 90 seconds and then annealed at 450 °C under high vacuum conditions to remove the contamination layers on the GaAs surface. The procedure followed to remove the contamination layer is as reported by Lebedev *et al.*[7] A commercially bought 5% Cu-doped ZnO target was used for sputtering with 1:1 ratio for the Ar:$O_2$ gases (base pressure was of the order of $10^{-6}$ Torr). The rf power during deposition was 75 W with a working pressure of 10 mTorr. Deposition was performed for 30 minutes and the substrate temperature was maintained at 400 °C. Some of the films so prepared were then annealed in air at 550 °C for 30 minutes. The As *K*-edge measurements were performed at the 10A beamline in fluorescence yield mode with an angle of 70° between the beam direction and the sample normal. XAS and XMCD measurements were performed in total electron yield mode at the 2A (Magnetic spectroscopy) beamline of Pohang Accelerator Laboratory (PAL), South Korea. The photon energy resolution was set at ~0.3 eV. For the XMCD measurements, the applied magnetic field was ~ 0.6 T and the degree of circular polarization of the beam was better than 95%. The magnetic properties were measured by the means of Quantum Design superconducting quantum interference device-vibrating sample magnetometer (SQUID-VSM). The hysteresis loop was performed by sweeping the magnetic field between -7 and 7 T.

Arsenic *K*-edge spectra can reveal the possible diffusion of As into ZnO films as the *K*-edge spectra are sensitive to the local environment of the probed ion. These spectra were collected for the as-deposited and annealed Cu-doped ZnO films and are shown in Fig. 1. Along with the As $K$-edge of GaAs substrate, as-deposited film and annealed film, the difference spectra are also plotted. The spectra of the substrate were subtracted from the spectra of the as-deposited and the annealed films. Before this subtraction, background was removed from all the spectra using the AUTOBK background removal algorithm employed in Athena software of the IFEFFIT suite of programs[8]. The overall shape and features of all the spectra look very similar. This is because the penetration depth of the hard X-rays employed in the measurements is much larger than the film thickness, so most of the probed ions contributing to the spectra are from the GaAs substrate. However, when we subtract the contribution from the substrate, we can clearly observe the signature from the films. From the difference spectra, the signatures of diffusion can be seen with the structures at 11861 and 11874 eV for the as-deposited film. For the annealed films, these features are shifted towards lower energy with their intensities increased. The changes observed are caused by the shift in peak position for the spectra of the films due to the change in the hybridization strength for As ions. Similar changes in the line-shape were also observed in the difference spectra (not shown here) between the as-deposited and the annealed films. Inset of Fig. 1 shows that there is small shift in the peak position of the As main peak towards higher energy. The atomic size of doped As ion (1.20 Å) is larger than the substituted O ion (0.73 Å). Thus in case of As-doped ZnO, larger hybridization is expected which can lead to a shift towards higher energy. The peak shifting and the increased intensity of the features in the difference spectra provide the signatures for the As-diffusion in Cu-doped ZnO films which is further enhanced by thermal annealing. Sun *et al*. deposited ZnO on GaAs using metal organic chemical vapor deposition (MOCVD) method[6]. They found that the as-deposited films were $n$-type and they could achieve $p$-type doping by annealing the films at 550 °C. The hole-doping concentration they measured was $1.45 \times 10^{18}$ cm$^{-3}$ and was assigned to the As diffusion from the GaAs substrate to the ZnO films. Following their conclusions, we suggest that in the present case also, the As diffusion may lead to hole doping. Further signature of the As diffusion-enhanced hole doping is obtained from the XAS measurements at the Cu $L_{3,2}$-edge measurements.

XAS spectra collected at transition metal $L_{3,2}$-edge is a sensitive probe for the charge state, local symmetry, hybridization and solid state effects of the probed ion. Figure 2 (a) shows the result of XAS measurements performed at 80 K. The XAS spectra of the as-grown and the annealed films are similar to that of Cu in 2+ valence state[9,10]. The $L_3$-edge contains mainly three features, marked by the labels $A$, $B$ and $C$ in Fig. 2 (a). $Cu^{2+}$ has a ground state which is predominantly $3d^9$ with a small configuration mixing of $3d^{10}\underline{L}$ where $\underline{L}$ denotes a hole in the oxygen ligand[9]. The possible excited states accessible (in accordance with the dipole selection rules) will be $\underline{2p}3d^{10}$, $\underline{2p}3d^94s^1$ and $\underline{2p}3d^{10}\underline{L}4s^1$ (here $\underline{2p}$ denotes a hole in the O $2p$ orbital). Following the assignment by van der Laan *et al.*, the peaks $A$ and $C$ are assigned to the transitions $3d^9 \rightarrow \underline{2p}3d^{10}$ and $3d^9 \rightarrow \underline{2p}3d^{10}\underline{L}4s^1$, respectively[9]. If there are holes present in the system due to the As diffusion, then the ground state for Cu $L_{3,2}$-edge will also consist of the configuration $3d^9\underline{L}$ ($Cu^{2+}$ plus an oxygen $p$ hole, formally $Cu^{3+}$). The peak $B$ at energy of 934.4 eV is an indication of the hole doping and is assigned to the transition $3d^9\underline{L} \rightarrow \underline{2p}3d^{10}\underline{L}$. A similar feature is reported by van der Laan *et al.*[9] in the $L_{3,2}$-edge of CuO at 935 eV due to the presence of defect state $3d^9\underline{L}$ in the ground state. This assignment is in accordance with other reports on high-$T_C$ Cu-based superconductors[10-12] and Cu-doped GaN nanowires[13], where a similar feature is reported to be an indication of the hole-doping in the system. Sarma *et al.*[10] argued the presence of holes in the high-$T_C$ superconductor $YBa_2Cu_7O_{7-d}$ based on the presence of a peak at 933 eV which appears as a shoulder to the main $L_3$-edge. Seong *et al.*, found similar structure at an energy of around 934.5 eV and attributed to the presence of $Cu^{3+}$-ion[13]. It has been used to quantitatively determine the $Cu^{3+}$/hole concentration in the system[12]. Following the procedure reported in the literature, peak *A* was fitted with a combination of Lorentzian and Gaussian functions and the small peak *B* with a Gaussian function. The average valence of Cu, $V(Cu)$, is calculated using the formula:

$$V(Cu) = 2 + \frac{I(B)}{I(L_3)}$$

Here $I(B)$ and $I(C)$ are the area under the feature *B* and the $L_3$-peak, respectively. The average valence of Cu for as-deposited and annealed films turns out to be 2.005 and 2.016. Using these numbers, the concentration of holes is calculated to be $1.05 \times 10^{19}$ and $3.36 \times 10^{19}$ cm$^{-3}$ for the as-deposited and the annealed films, respectively. The values for the hole concentration calculated using this method is an order of magnitude larger than the values reported by Sun *et al.*[6] which may be due to the hole contribution from the doped Cu ions. Although these values may not be very accurate, however, they do provide evidence for the hole-doping in the system with increase in hole concentration upon thermal annealing.

XMCD spectra can be used to investigate the exchange splitting of the electronic states of the probed ion at the Fermi level and hence giving information on the ion-specific magnetic contribution. Dipole selection rules allow us to measure the magnetic contribution from the 3*d* orbitals of the doped-Cu ions from the XMCD measurements at the Cu $L_{3,2}$ – edge. Results of the XMCD measurements performed at 80 K for the as-deposited and annealed film are shown in of Fig. 2 (b). From the XMCD spectra we can clearly observe the ferromagnetic signal arising from the doped Cu ion's $3d^9$ orbital. The XMCD signal for the annealed film is twice compared to the as-deposited film. With the help of results discussed previously, we have shown enhanced As-diffusion and hole doping with thermal annealing of the film. Thus, the enhancement in the XMCD signal for the annealed film supports the idea that the ferromagnetism observed in the present case is hole-mediated and can be strengthened with introduction of more holes in the system. Based on these results, it is proposed that the exchange interaction among the localized moments of doped Cu ions is mediated by the holes in the system.

Figure 3 show the result of the SQUID measurements performed at 2 K, 80 K and 300 K for as-deposited and annealed film. From the figure it can be seen that the ferromagnetic signal from the annealed film is twice as large as that for the as-deposited film at 80 K which is well in accordance with the XMCD results. Fig. 3(b) shows an enlarged view of the measurement for the annealed film at 2 K and 80 K. Results for all the films show that the magnetization values saturate for low magnetic fields and that the films are ferromagnetic even at room temperature. The magnetic signal of the films at 300 K has values slightly reduced in comparison to the values at 80 K. At 2 K, there is an increase in the magnetization of the films, with as-deposited and annealed films showing almost similar behavior. The hole concentration in the present case is found to be an order of magnitude less than that reported by Dietl *et al.* for above room temperature ferromagnetism. Due to the less number of charge carriers, the exchange interaction among all the local magnetic moments is expected to be less. Thus, not all the doped Cu ions are ferromagnetically aligned. Number of magnetically aligned ions is increasing with decrease in temperature resulting in the observed increase in magnetization values.

We have successfully prepared ZnO:Cu films using rf-sputtering technique on GaAs substrate which are ferromagnetic at room temperature. Films are doped *p* –type due to As diffusion from the substrate to the film enhanced by thermal annealing. The magnetism in the films is hole-mediated and is shown to strengthen with the increase in the hole concentration due to the As diffusion. Arsenic diffusion is indicated by the As $K$-edge measurements. Hole doping due to the As diffusion is evidenced by the XAS measurements at Cu $L_{3,2}$-edge. XMCD and magnetic measurements show that the films are ferromagnetic and provide support to the idea of hole-mediated ferromagnetism in the present study as suggested by Dietl *et al.*[2]

This work was supported by NRF under Contract Nos. 2009-0088969 and 2010-0018733. The experiment at PLS is supported by POSTECH and MEST. M.H.J. acknowledges the support by the NRF (2010-0005427).

# References

[1] D. D. Awschalom and M. E. Flatté, Nat. Phys. **3**, 153 (2007).

[2] T. Dietl, H. Ohno, F. Matsukura, J. Cibert, and D. Ferrand, Science **287**, 1019 (2000).

[3] T. Herng, D. C. Qi, T. Berlijn, J. Yi, K. Yang, Y. Dai, Y. Feng, I. Santoso, C. Sánchez-Hanke, X. Gao, Andrew Wee, W. Ku, J. Ding, and A. Rusydi, Phys. Rev. Lett. **105** (2010).

[4] T. Dietl, A. Haury, and Y. Merle d'Aubigné, Phys. Rev. B **55**, R3347 (1997).

[5] Y. R. Ryu, S. Zhu, D. C. Look, J. M. Wrobel, H. M. Jeong, and H. W. White, J. Cryst. Growth **216**, 330 (2000); D. C. Look, G. M. Renlund, R. H. Burgener, and J. R. Sizelove, Appl. Phys. Lett. **85**, 5269 (2004).

[6] J. C. Sun, J. Z. Zhao, H. W. Liang, J. M. Bian, L. Z. Hu, H. Q. Zhang, X. P. Liang, W. F. Liu, and G. T. Du, Appl. Phys. Lett. **90**, 121128 (2007).

[7] M. Lebedev, Appl. Surf. Sci. **229**, 226 (2004).

[8] B. Ravel and M. Newville, J. Synchrotron. Radiat. **12**, 537 (2005).

[9] G. van der Laan, R. A. D. Pattrick, C. M. B. Henderson, and D. J. Vaughan, J. Phys. Chem. Solids **53**, 1185 (1992).

[10] D. D. Sarma, O. Strebel, C. T. Simmons, U. Neukirch, G. Kaindl, R. Hoppe, uuml, and H. P. ller, Phys. Rev. B **37**, 9784 (1988).

[11] C. F. J. Flipse, G. van der Laan, A. L. Johnson, and K. Kadowaki, Phys. Rev. B **42**, 1997 (1990).

[12] M. Karppinen, M. Kotiranta, T. Nakane, H. Yamauchi, S. C. Chang, R. S. Liu, and J. M. Chen, Phys. Rev. B **67**, 134522 (2003); M. Karppinen, S. Lee, J. M. Lee, J. Poulsen, T. Nomura, S. Tajima, J. M. Chen, R. S. Liu, and H. Yamauchi, Phys. Rev. B **68**, 054502 (2003).

[13] Han-Kyu Seong, Jae-Young Kim, Ju-Jin Kim, Seung-Cheol Lee, So-Ra Kim, Ungkil Kim, Tae-Eon Park, and Heon-Jin Choi, Nano Lett. **7**, 3366 (2007).

# Figure Captions

**FIG. 1.** (color online) As $K$-edge spectra for the GaAs substrate, as-deposited and annealed Cu-doped ZnO thin films (solid lines). Difference spectra between the films and the substrate are plotted in dotted lines.

**FIG. 2.** (color online) Cu $L_{3,2}$-edge (a) XAS and (b) XMCD spectra for the as-deposited and annealed Cu-doped ZnO thin films measured at 80 K.

**FIG. 3.** (color online) M-H loop measurements performed at (a) 2 K, (c) 80 K and (d) 300 K for the as-deposited and annealed films. Fig. 3 (b) shows an enlarged view of measurements performed at 2 K and 80 K for the annealed film.

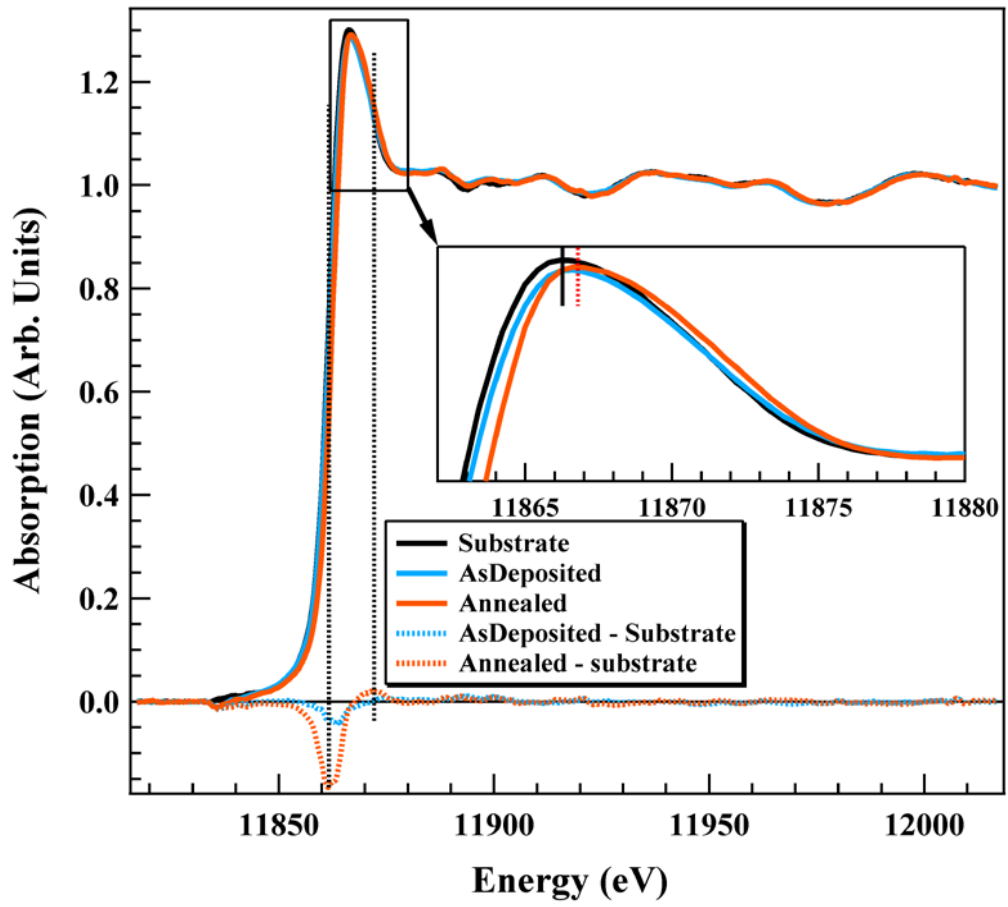

**FIG. 1** (Abhinav Pratap Singh *et al.*)

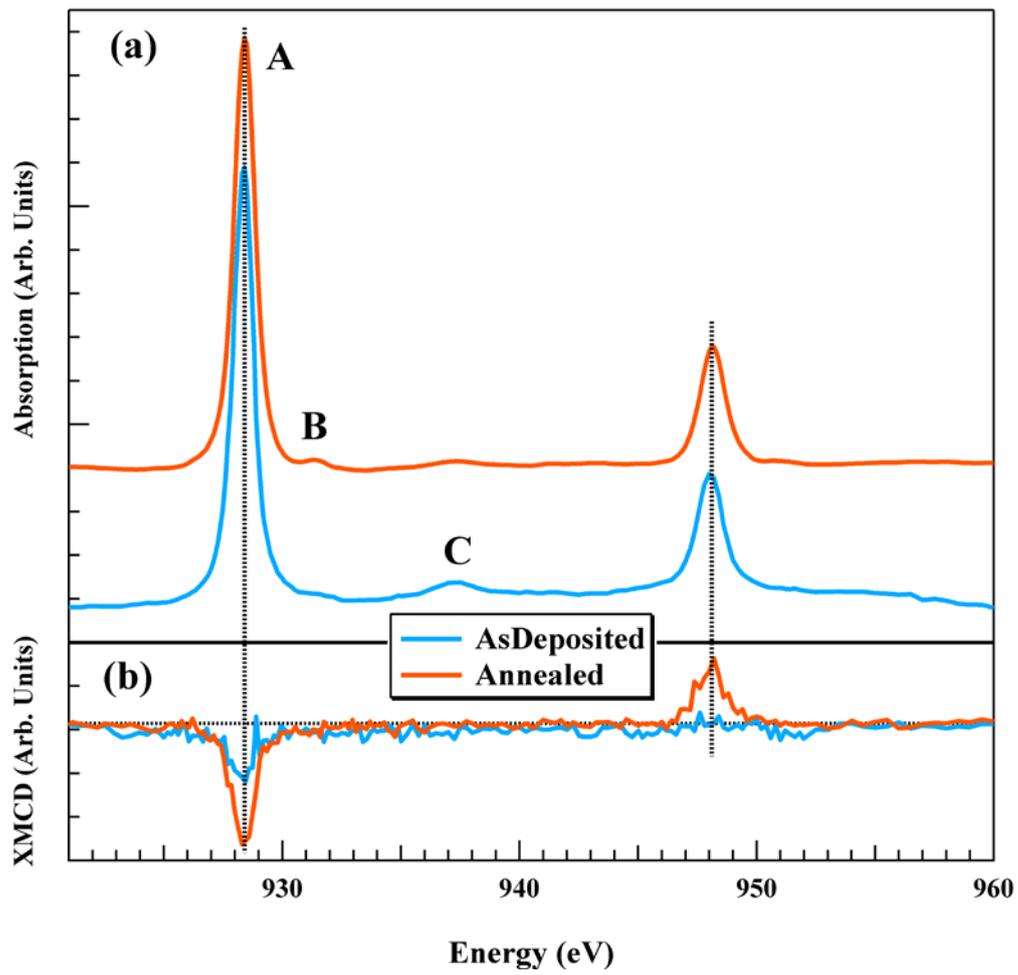

**FIG. 2** (Abhinav Pratap Singh *et al.*)

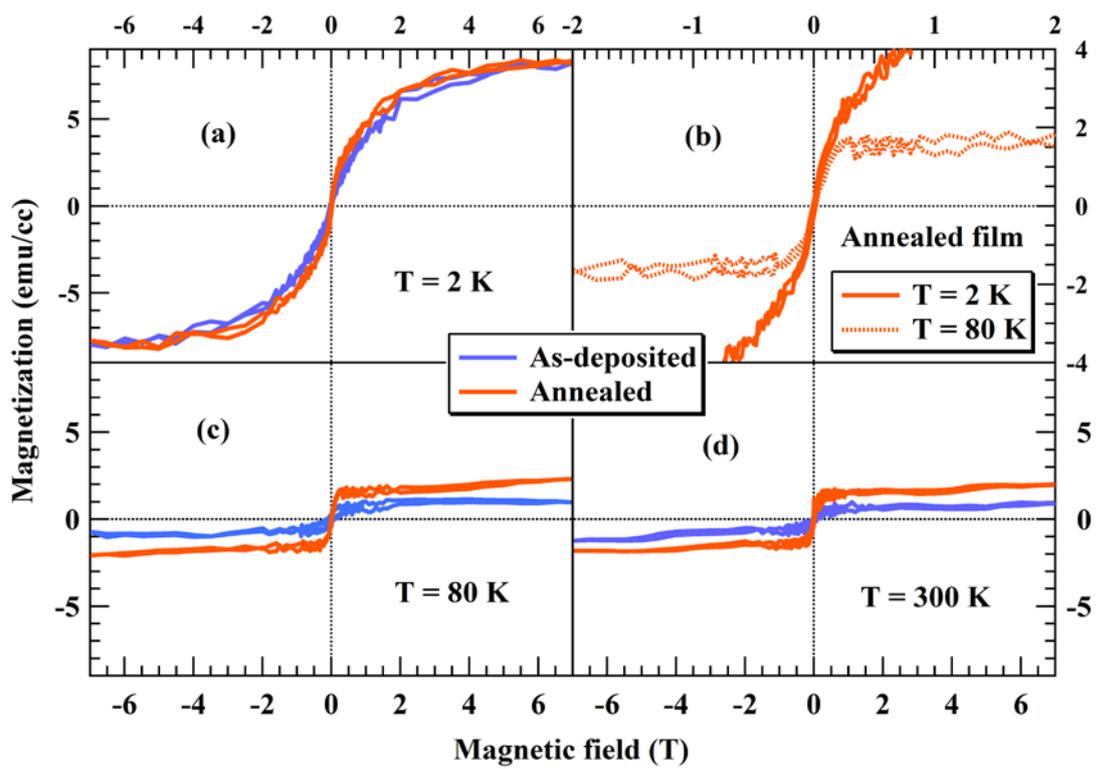

**FIG. 3** (Abhinav Pratap Singh *et al.*)